%% file: paper.tex
\documentclass[12pt,a4paper]{iopart}

\usepackage{aas_macros}

\usepackage{url}
\usepackage{graphicx}
\usepackage{booktabs}
\usepackage{makecell}
\usepackage[separate-uncertainty=true]{siunitx}
\DeclareSIUnit{\sqrthz}{\ensuremath{\sqrt{\text{\hertz}}}}
\DeclareSIUnit{\vrms}{\volt_{RMS}}
\DeclareSIUnit{\mps}{(ms^{-1})}

\usepackage{natbib}

\begin{document}

\title[Atmospheric scintillation during a period of Calima at Observatorio del Teide]
    {Measurement of atmospheric scintillation during a period of Saharan dust (Calima) at Observatorio del Teide, I{\~z}ana, Tenerife, and the impact on photometric exposure times}

\author{S J Hale$^{1,2}$, W J Chaplin$^{1,2}$, G R Davies$^{1,2}$, Y P Elsworth$^{1,2}$, R Howe$^{1,2}$, P L {Pall{\'e}}$^{3}$}
\address{$^1$ School of Physics and Astronomy, University of
  Birmingham, Edgbaston, Birmingham B15 2TT, United Kingdom}
\address{$^2$ Stellar Astrophysics Centre, Department of Physics and
  Astronomy, Aarhus University, Ny Munkegade 120, DK-8000 Aarhus C,
  Denmark}
\address{$^3$ Instituto de Astrof{\'i}sica de Canarias, and Department
  of Astrophysics, Universidad de La Laguna, San Crist{\'o}bal de La
  Laguna, Tenerife, Spain}
\ead{s.j.hale@bham.ac.uk}

\begin{indented}
\item[]{}\quad\rm\ignorespaces

This is an author-created, un-copyedited version of an article
accepted for publication in Publications of the Astronomical Society
of the Pacific. The publisher is not responsible for any errors or
omissions in this version of the manuscript or any version derived
from it.
\end{indented}

\begin{abstract}
\input{abstract}
\end{abstract}


\maketitle

\ioptwocol

\input{introduction}
\input{conditions}
\input{scintillation}
\input{results}
\input{discussion}
\input{opendata}

\ack

\input{acknowledgements}


\bibliographystyle{plainnat}
\bibliography{references}

\end{document}

%% file: abstract.tex
%
%
%


We present scintillation noise profiles captured at the Observatorio
del Teide, Iza{\~n}a, Tenerife, over a one-week period in
September~2017.  Contemporaneous data from the Birmingham Solar
Oscillations Network (BiSON) and the Stellar Activity (STELLA) robotic
telescopes provides estimates of daily atmospheric extinction allowing
the scintillation noise to be placed within the context of overall
atmospheric conditions.  We discuss the results both in terms of the
impact on BiSON spectrophotometer design, and for astronomical
observations more generally.  We find that scintillation noise power
reduces by half at about~\SI{5}{\hertz}, and is reduced to one tenth
between~\SIrange{20}{30}{\hertz} even during periods of mild Calima,
where visibility is reduced due to high concentrations of mineral dust
in the atmosphere.  We show that the common accepted exposure time
of~\SI{<10}{\milli\second} for limiting the effect of scintillation
noise in ground based photometry may be increased, and that depending
on the application there may be little benefit to achieving exposure
times shorter than~\SI{50}{\milli\second}, relaxing constraints on
detector gain and bandwidth.


%% file: introduction.tex
%
%
%
%


\section{Introduction}

The Birmingham Solar Oscillations Network (BiSON) is a six-site
ground-based network of spectrophotometers observing oscillations of
the Sun~\citep{s11207-015-0810-0}.  For many ground-based photometers,
the dominant noise source is that from atmospheric
scintillation~\citep{2015MNRAS.452.1707O}.  The BiSON
spectrophotometers seek to reduce the effect of atmospheric
scintillation by making use of a multi-wavelength polarisation
switching technique.  The basic measurement is Doppler velocity,
determined from the changes in intensity at two points in the wings of
the potassium absorption line at~\SI{770}{\nano\metre}.  Two short
(\SI{\approx5}{\milli\second}) exposures are taken consecutively and
the ratio, $R$, of intensity between the two wings is determined,
\begin{equation}
  R = \frac{I_{\mathrm{b}} - I_{\mathrm{r}}}{I_{\mathrm{b}} + I_{\mathrm{r}}} ~,
\end{equation}
where $I_{\mathrm{b}}$ and $I_{\mathrm{r}}$ are the intensities
measured at the blue and red wings of the solar absorption line,
respectively.  The ratio is proportional to the observed line-of-sight
Doppler velocity shift.  Integrating many short exposures and
calculating the ratio in this way allows common intensity fluctuations
to cancel, reducing the effect of atmospheric scintillation noise.

The BiSON node located at Iza{\~n}a on the island of Tenerife, at the
Instituto de Astrof{\'i}sica de Canarias
(IAC)~\citep{2014MNRAS.443.1837R}, suffers additional complications in
terms of visibility and atmospheric noise due to the proximity of the
Western Sahara, just~\SI{100}{\kilo\metre} from the North African
coast.  During the summer months the Canary Islands frequently
experience high concentrations of mineral dust in the atmosphere,
known as Calima, where the Saharan Air Layer passes over causing a
fog-like reduction in visibility and a change in atmospheric
characteristics.  The aim of this paper is to determine the
contribution of atmospheric scintillation noise to the overall noise
budget in BiSON observations, and to investigate the change in
atmospheric noise characteristics with periods of Calima.  There are
several techniques for estimation and approximation of the effect of
scintillation noise -- see, e.g.,
\citet{1967AJ.....72..747Y,1969ApOpt...8..869Y,1997PASP..109..173D,1998PASP..110..610D,2006PASP..118..924K,2012A&A...546A..41K,SHEN2014160,2019MNRAS.489.5098F}.
Here, we take an empirical approach to exploring the temporal
frequency spectrum of the scintillation.  The scintillation noise
measurements presented will be of general interest to astronomers at
Iza{\~n}a, however it should be noted that these day-time results will
be worse than night-time conditions.

In Section~\ref{conditions} we look at two methods of estimating
atmospheric conditions, and in Sections~\ref{data} and~\ref{results}
we present scintillation noise data captured over several days in
September~2017 during a period of mild Calima.  Finally, in
Section~\ref{discussion} we discuss the results both for the impacts
on BiSON spectrophotometer design, and for astronomical observations
more generally.


%% file: conditions.tex
%
%
%
%


\section{Atmospheric conditions}
\label{conditions}

\begin{figure*}[t]
  \centering
  \includegraphics[scale=1.0]{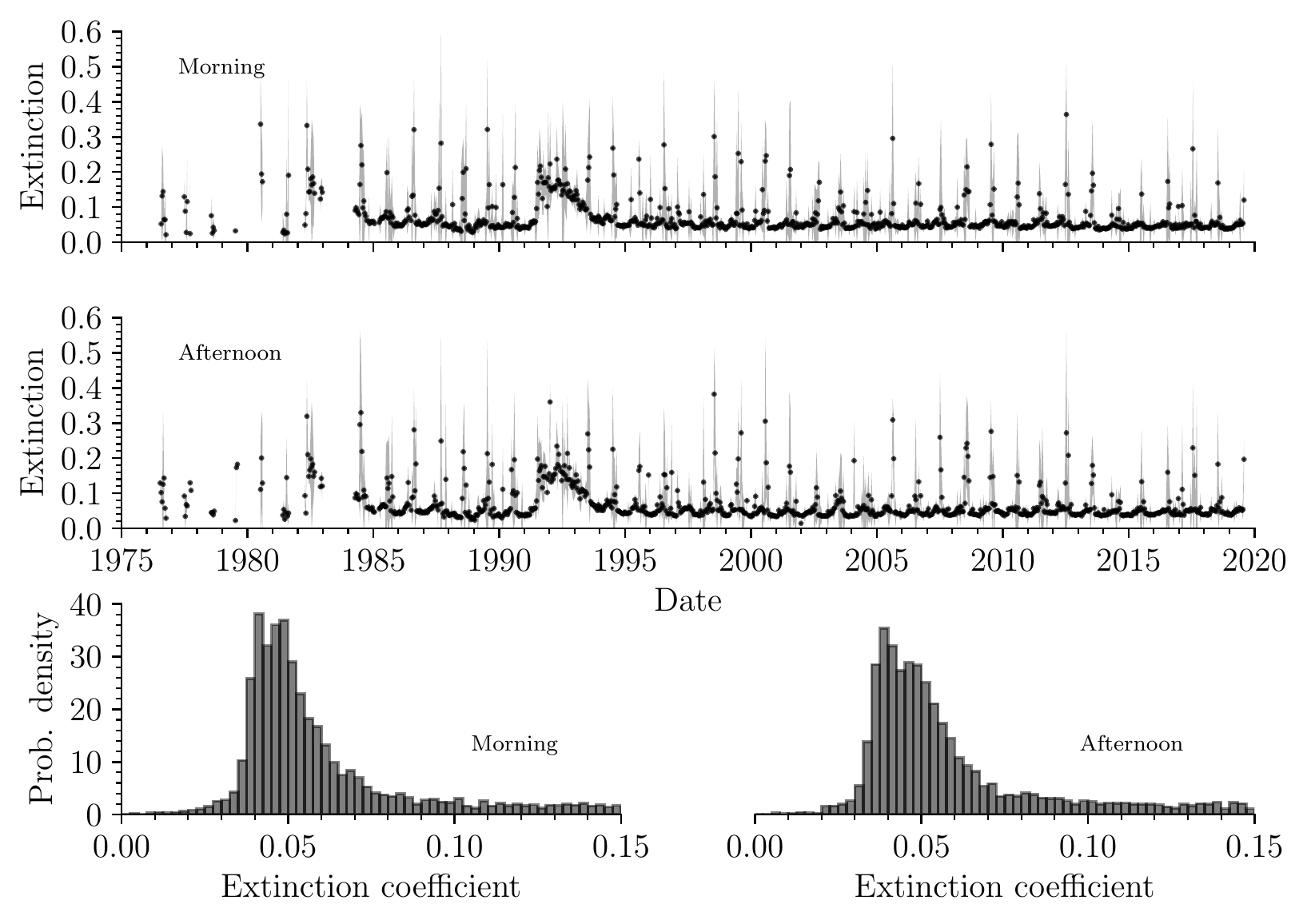}
  \caption{Extinction coefficients and statistical distribution from
    Iza{\~n}a, Tenerife, estimated with the Mark-I BiSON instrument
    using the technique described by~\citet{1538-3881-154-3-89}.
    Morning and afternoon extinction values are estimated separately,
    shown in the top and middle panel.  Each dot represents the median
    value over 14~days.  The grey shading represents $\pm3$ times the
    standard-error on each median value.  The large increase in
    extinction during the period between 1991 and 1994 was caused by
    the eruption of the Mount Pinatubo volcano in the Philippines.
    The bottom two panels show the statistical distribution of
    extinction, which is Gaussian with a long-tail caused by the
    effects of Calima.}
  \label{fig:izana_extinction}
\end{figure*}

In order to interpret the changes in atmospheric scintillation noise
characteristics it is essential to also estimate the overall
atmospheric conditions. This is particularly relevant to Iza{\~n}a,
where placing results on scintillation in context requires an
understanding of the impact of the Calima.  The atmospheric extinction
coefficient was determined each day using the technique described
by~\citet{1538-3881-154-3-89}.  In brief, this involved logging the
changing solar intensity measured throughout the day by our BiSON
instrument, and fitting this for both morning and afternoon periods
against the airmass calculated from the known solar zenith angle,
\begin{equation}
  \ln(I/I_0) = -\tau A ~,
\end{equation}
where $I$ is the direct-Sun radiance, $I_0$ the maximum intensity
measured on a given day, and $A$ the airmass.  The gradient of the
fit, $\tau$, is a measure of the column atmospheric aerosol optical
depth (AOD) per unit airmass, calibrated in terms of
magnitudes~per~airmass.  Figure~\ref{fig:izana_extinction} shows the
estimated atmospheric extinction coefficients determined from the
archive of data from the Mark-I BiSON instrument at Iza{\~n}a.  The
typical extinction coefficient at Iza{\~n}a on a clear day is
about~\num{0.05}~magnitudes~per~airmass.  During mineral dust events,
often between June and October, extinction can rise to values
between~\num{0.1} and~\num{0.8}~magnitudes~per~airmass.

\begin{figure*}
  \centering
  \includegraphics[scale=1.0]{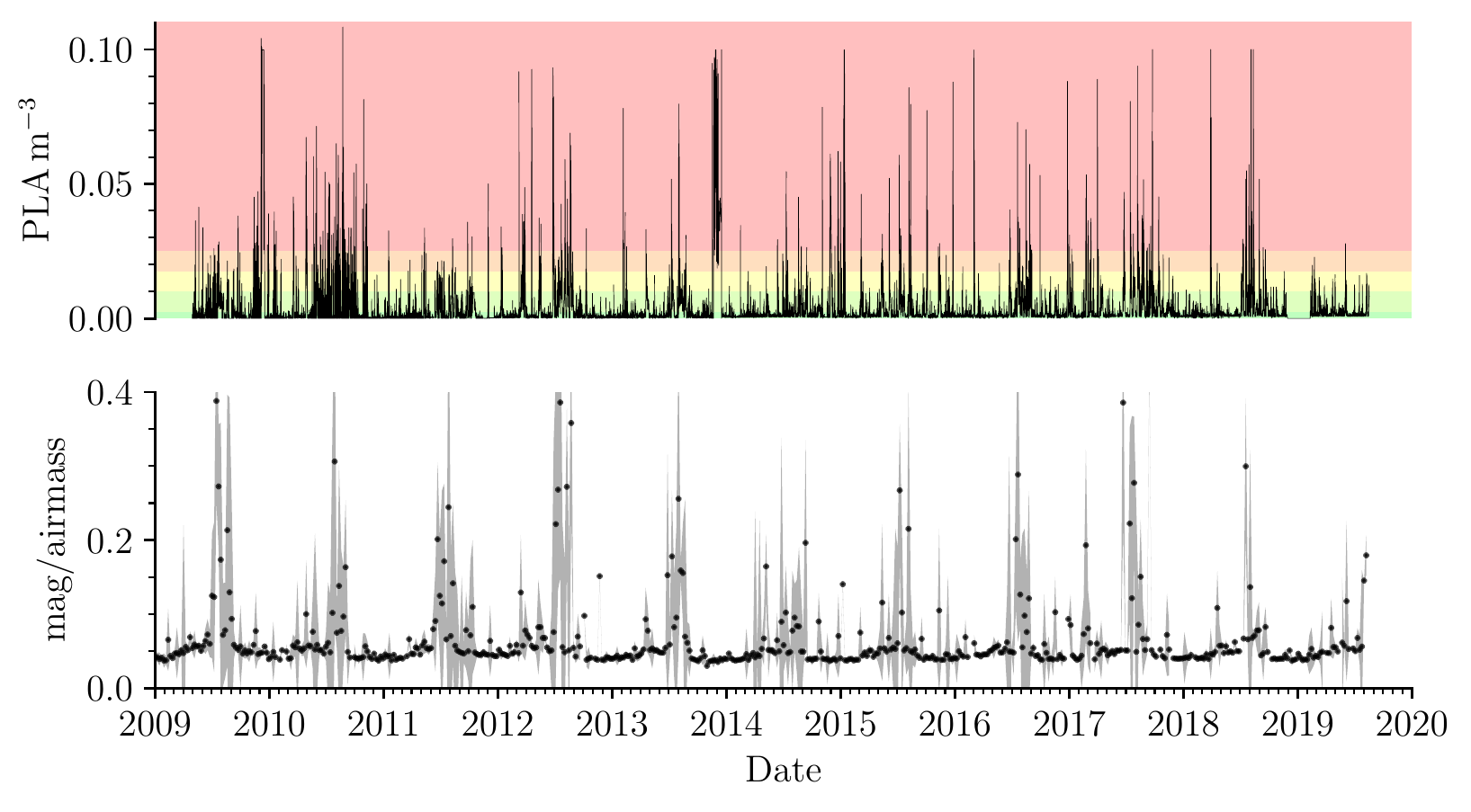}
  \caption{Top: Dust data obtained with the~\citet{dust} at Iza{\~n}a,
    an AIP facility jointly operated by AIP and IAC.  The coloured
    bands indicate the qualitative limits described in
    Table~\ref{table:dust_limits}.  Towards the end of 2018 a fault
    was discovered with the wiring of the sensor, which following the
    repair resulted in a cleaner signal and calibration change.
    Bottom: Average daily extinction coefficients at Iza{\~n}a
    estimated with the BiSON instrument over the same time period.
    Each dot represents the median value over 7~days.  The grey
    shading represents $\pm3$ times the standard-error on each median
    value.  The Spearman rank-order correlation coefficient for the
    two datasets is~\num{0.5} with an infinitesimally small
    false-alarm probability.}
  \label{fig:dust}
\end{figure*}

A real-time estimate of ground-level dust at Iza{\~n}a is provided by
the Stellar Activity (STELLA) robotic telescopes~\citeyearpar{dust}, a
Leibniz Institute for Astrophysics Potsdam (AIP) facility. STELLA make
available, amongst other parameters, measurements from a VisGuard~2
In-situ Visibility Monitor manufactured by~\citet{14289E/4}.  The
VisGuard is a photometer that measures the intensity of scattered
light from an air sample drawn into the instrument by a fan.  The
output is a measure of Polystyrol-Latex-Aerosols (PLA) per cubic
metre, and when the instrument is installed in its intended use case
monitoring visibility in vehicle tunnels, this value can be converted
to an extinction coefficient.  When used in open-air observations, the
factory calibration is uncertain.  The IAC have therefore defined some
threshold levels for what is considered to be low and high, and these
are detailed in Table~\ref{table:dust_limits}.  The archive of data
from the VisGuard is shown in Figure~\ref{fig:dust} with coloured
banding to indicate the IAC thresholds.  The period of unusually high
values towards the end of 2013 results from a sensor glitch, and at
the end of 2018 a repair to the wiring of the sensor has reduced the
noise level and resulted in a change of calibration.

\begin{table}
    \caption{Qualitative limits used at the IAC in calibration of
      ground-level dust measurements from the~\citet{dust}.}
    \label{table:dust_limits}
    \footnotesize
    \centering
    \begin{tabular}{l l c c}
      \toprule

      \multicolumn{2}{c}{\makecell[cb]{Qualitative Value}} &
      \multicolumn{2}{c}{\makecell[cb]{Range ($\mathrm{PLA\,m^{-3}}$)}}
      \\
      \cmidrule(lr){1-2}\cmidrule(lr){3-4}
      Spanish & English & Min & Max\\
      
      \midrule

      Poca          & Little         &              & \num{<0.0025}\\
      Media         & Medium         & \num{0.0025} & \num{<0.0100}\\
      Bastante      & Quite a lot    & \num{0.0100} & \num{<0.0175}\\
      Mucha         & A lot          & \num{0.0175} & \num{<0.0250}\\
      Fuera limites & Outside limits & \num{0.0250} &\\

      \bottomrule

    \end{tabular}
\end{table}

The two results do not necessarily show perfect correlation, since the
BiSON atmospheric extinction measurement is estimating the column
aerosol optical depth through the entire atmosphere, and the VisGuard
is estimating the amount of dust at only ground level.  In general,
when the extinction is high the PLA tends to be high, and the Spearman
rank-order correlation coefficient is~\num{0.5} with an
infinitesimally small false-alarm probability.  When considered
together the two measurements provide a good qualitative judgement of
atmospheric conditions, and this allows us to better understand the
following frequency-dependent scintillation measurements, which show
significant variation.


%% file: scintillation.tex
%
%
%
%


\section{Scintillation measurement}
\label{data}

\begin{figure*}[p]
  \centering
  \includegraphics[scale=1.0]{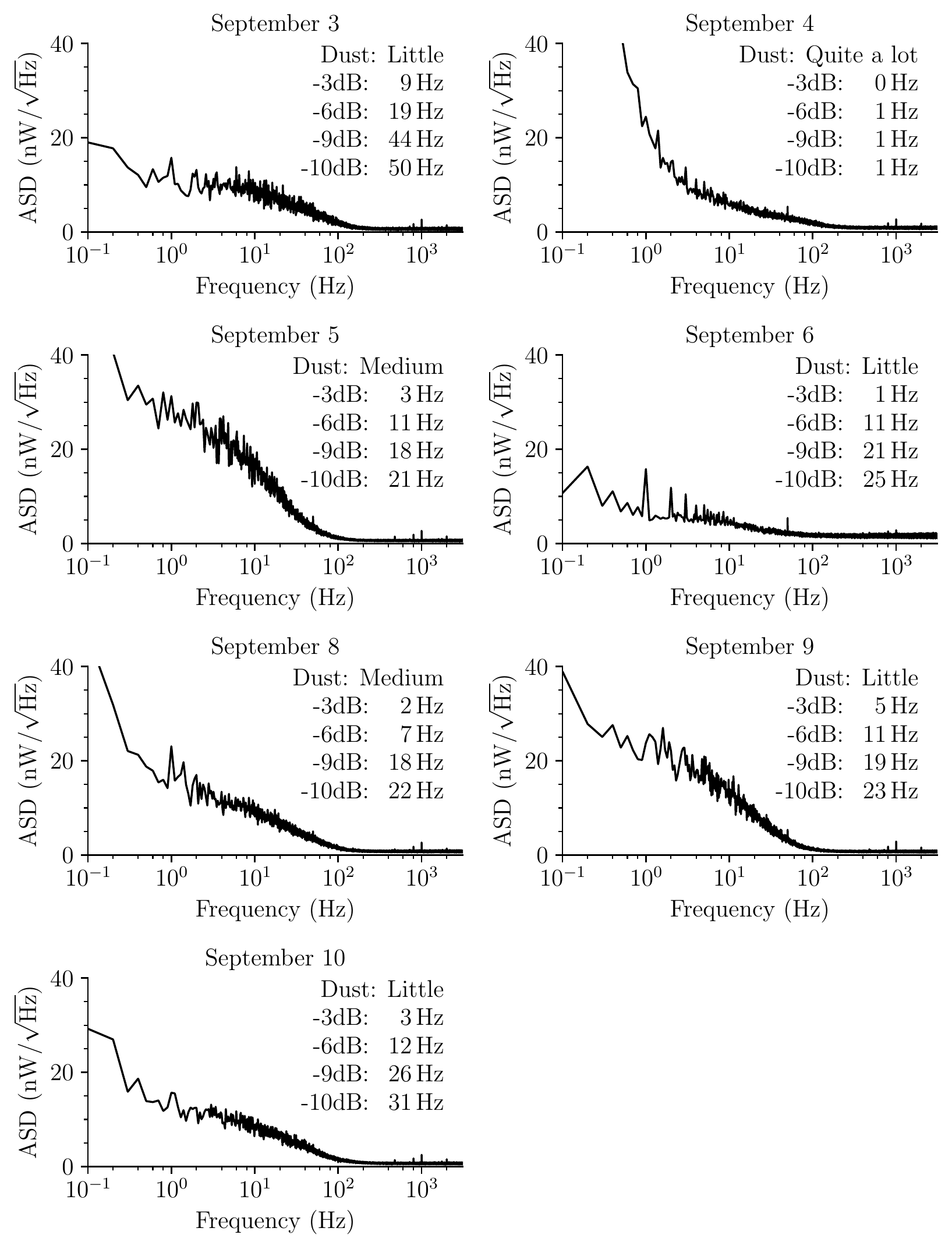}
  \caption{Daily scintillation measurements collected during
    2017~September, using the technique described in
    Section~\ref{data}. The qualitative dust levels are described in
    Table~\ref{table:dust_limits}.  The absolute values for
    September~6 are low due to a temporary change in collection
    optics.}
\label{fig:scintillation}
\end{figure*}

\begin{table*}
    \sisetup{table-number-alignment=center,separate-uncertainty=true,table-format=2.0,round-mode=places,round-precision=0}
    \caption{Day-time scintillation noise characteristics and atmospheric conditions, including telescope operator notes on night-time observations.}
    \label{table:noise_characteristics}
    \centering
    \begin{tabular}{l S[table-format=2.1,round-precision=1] S S S S S[table-format=1.4,round-precision=4] S[table-format=1.3(1),round-precision=3]}

    \toprule

    {Date} & {Mean} &
    \SI{-3}{\decibel} & \SI{-6}{\decibel} & \SI{-9}{\decibel} & \SI{-10}{\decibel} & {Dust} & {Extinction}\\

    & {($\mathrm{nW}/\sqrt{\mathrm{Hz}}$)} & {(\si{\hertz})} & {(\si{\hertz})} & {(\si{\hertz})} & {(\si{\hertz})} & {($\mathrm{PLA\,m^{-3}}$)} & {(mag/airmass)}\\

    \midrule

    2017/09/03 & 12.269335254522133 & 9.0999999999909 & 19.4999999999805 & 43.79999999995619 & 50.199999999949796 & 0.0017 & 0.044(4)\\[-1ex]
    \multicolumn{8}{l}{\tiny Not photometric, thin cirrus.}\\[+0.5ex]
    2017/09/04 & 47.265457448820186 & 0.39999999999959995 & 0.5999999999993999 & 1.0999999999988999 & 1.3999999999986 & 0.0150 & 0.111(14)\\[-1ex]
    \multicolumn{8}{l}{\tiny 100\% clear. Variable night, with rising calima and almost full moon. Loose wind and medium humidity. Photometric except for the moon.}\\[+0.5ex]
    2017/09/05 & 30.422448602274272 & 2.7999999999972 & 10.999999999988999 & 18.3999999999816 & 21.2999999999787 & 0.0080 & 0.322(32)\\[-1ex]
    \multicolumn{8}{l}{\tiny 100\% clear. Almost no calima and full moon. Moderate wind and very low humidity. Photometric except for the moon.}\\[+0.5ex]
    2017/09/06 & 8.530052530037192 $^{\dagger}$ & 1.0999999999988999 & 10.999999999988999 & 20.699999999979298 & 25.099999999974898 & 0.0009 & 0.048(4)\\[-1ex]
    \multicolumn{8}{l}{\tiny 100\% clear. No calima and almost full moon. Moderate wind and very low humidity. Photometric except for the moon.}\\[+0.5ex]
    2017/09/07 & & & & & & 0.0100 & 0.046(4)\\[-1ex]
    \multicolumn{8}{l}{\tiny 100\% clear. Excellent seeing. Photometric except for the moon.}\\[+0.5ex]
    2017/09/08 & 19.69368160072787  & 1.5999999999983998 & 7.399999999992599 & 17.999999999982 & 22.1999999999778 & 0.0025 & 0.222(20)\\[-1ex]
    \multicolumn{8}{l}{\tiny Not photometric, thin cirrus. Excellent seeing.}\\[+0.5ex]
    2017/09/09 & 23.99553882893885  & 5.199999999994799 & 11.2999999999887 & 18.999999999980997 & 22.899999999977098 & 0.0015 & 0.172(23)\\[-1ex]
    \multicolumn{8}{l}{\tiny Not photometric, thin cirrus.}\\[+0.5ex]
    2017/09/10 & 15.414967779700888 & 3.2999999999967 & 11.999999999987999 & 25.999999999973998 & 31.399999999968596 & 0.0017 & 0.096(8)\\[-1ex]
    \multicolumn{8}{l}{\tiny 100\% clear. Photometric except for the moon.}\\









    \bottomrule

    \end{tabular}
    \flushleft \footnotesize $^{\dagger}$This value is low due to a temporary change in collection optics.
\end{table*}

The scintillation noise was measured at about the same time each
morning over a one-week span in~2017.  This period captured several
days of marked variation in sky quality.  Independent measures of sky
quality were available on each day from BiSON (extinction) and STELLA
(dust PLA).

Sunlight was collected using the Solar Pyramid coelostat at the
Observatorio del Teide, and a small bespoke instrument (independent of
the BiSON spectrophotometer) was used to log intensity fluctuations.
The instrument consisted of a~\SI{25.4}{\milli\metre} objective lens,
filtered to a bandwidth of~\SIrange{700}{900}{\nano\metre}, with
sunlight fed through an optical fibre to a simple photodiode and
transimpedance amplifier.  The wavelength bandwidth was selected to
match existing BiSON instrumentation.  The input power with this
configuration was approximately~\SI{0.05}{\milli\watt}, producing an
output of approximately~\SI{700}{\milli\volt} after amplification with
a~\SI{30}{\kilo\ohm} gain resistor and~\SI{0.45}{\ampere\per\watt}
photodiode quantum efficiency.  The detector bandwidth was
approximately~\SI{50}{\kilo\hertz}.  The light intensity was logged
using a digital oscilloscope for~\SI{10}{\second}
at~\SI{6.25}{\kilo\hertz}.  A total of~\num{24} realisations of noise
were captured consecutively each morning, producing four minutes of
data in~\SI{10}{\second} segments.  The spectral density of
each~\SI{10}{\second} segment was calculated and then stacked to
improve the final atmospheric ``signal'' to noise ratio.


%% file: results.tex
%
%
%
%


\section{Results}
\label{results}

The results from the scintillation measurements are shown in
Figure~\ref{fig:scintillation}, where each panel shows the amplitude
spectral density (ASD) of scintillation noise measured each day, with
annotations indicating dust level and frequencies of power roll-off.
The effects of Calima are most pronounced below~\SI{5}{\hertz},
equivalent to exposures of~\SI{\geq200}{\milli\second}.  Photometry
techniques making use of multi-wavelength observations to reduce the
effect of scintillation noise typically require high-speed
(\SI{<10}{\milli\second}) exposures~\citep{1997PASP..109..173D}, and
we can see from the profiles in Figure~\ref{fig:scintillation} that
the effect of scintillation noise has indeed almost completely decayed
at frequencies above~\SI{100}{\hertz}, even during days of medium to
high Calima.  The mean value of the ``white'' band of scintillation
noise between~\SIrange{0.5}{1.0}{\hertz} was measured, and this value
was used as the baseline to determine the frequencies of
the~\SIrange{-3}{-10}{\decibel} points, listed in
Table~\ref{table:noise_characteristics}.  For comparison, the IAC
telescope operator night-time logs are also included for each day.
Exposures shorter than~\SI{200}{\milli\second}, equivalent
to~\SI{5}{\hertz}, see the scintillation noise power reduced by half
(\SI{-3}{\decibel}) over that of longer exposures.  A reduction in
scintillation noise of~\SI{-6}{\decibel} is typically achieved at
about~\SI{11}{\hertz}, and reduction of~\SI{-10}{\decibel}
between~\SIrange{20}{30}{\hertz}, which means depending on the
application there may be little benefit to achieving exposure times
shorter than~\SI{50}{\milli\second}, relaxing constraints on detector
gain and bandwidth.  It is expected that results will differ between
day- and night-time observations, and for instruments with different
aperture size.


%% file: discussion.tex
%
%
%
%


\section{Discussion}
\label{discussion}

BiSON spectrophotometers have made use of various optoelectronic
components to achieve computer-controlled wavelength selection via
polarisation switching, with rates varying from~\SI{0.5}{\hertz}
to~\SI{50}{\kilo\hertz}.  Modern BiSON spectrophotometers use Pockels
effect cells switched at~\SI{90.9}{\hertz}.  The data acquisition
subsystem takes~\SI{5}{\milli\second} exposures per polarisation
phase, plus~\SI{0.5}{\milli\second} stabilisation time at each
switching point, making an~\SI{11}{\milli\second} period.  This is
repeated 340~times for~\SI{3.74}{\second}, giving a total exposure
integration of~\SI{3.4}{\second} within each~\SI{4}{\second} sample.
Faster switching has the detrimental effect of reducing the overall
integration time due the need for stabilisation periods at each
switching interval.  Slower switching is less effective at removal of
scintillation noise but reduces the dead time.  In order to determine
the optimum switching rate it is necessary to know the characteristics
of atmospheric scintillation noise.

Many polarisation switching techniques are not capable of reaching
rates as high as~\SI{100}{\hertz}, and so during instrumentation
design it is essential to know the potential impact of atmospheric
scintillation on the overall noise levels, and how this compares with
other noise sources in the system.  Electronic noise, photon shot
noise, and noise from thermal fluctuations each contribute less
than~\SI{1}{\pico\watt\per\sqrthz} to the overall noise level, and by
comparison with the values shown in
Table~\ref{table:noise_characteristics}, we see the system is
dominated by scintillation noise as is expected for ground-based
photometric measurements.

\begin{table}
    \centering
    \caption{Velocity-calibrated white noise level and noise
      equivalent velocity (NEV) for five BiSON spectrophotometers over
      a 2018 summer observing campaign.}
    \label{table:2018_performance}

    \begin{tabular}{l S[table-format=4.1] S[table-format=4.1]}

    \toprule

    Site & {White noise}               & {NEV}\\
         & \si{\mps\squared\per\hertz} & \si{\centi\meter\per\second}~RMS\\

    \midrule

             Sutherland &  4.3 & 23.1\\ 
      \textbf{BiSON:NG} &  9.9 & 35.2\\
               Narrabri & 10.4 & 36.1\\
           Las~Campanas & 13.5 & 41.1\\
                 Mark-I & 61.6 & 87.7\\

    \bottomrule

    \end{tabular}

\end{table}

Throughout 2018, a next generation prototype BiSON spectrophotometer
was commissioned at Iza{\~n}a alongside the Mark-I instrument, sharing
light from the pyramid coelostat~\citep{halephd}.  The aim of
{BiSON:NG} is to miniaturise and simplify the instrumentation as much
as practical, typically though the use of off-the-shelf components,
whilst maintaining performance comparable with the existing network.
The prototype spectrophotometer makes use of an LCD retarder for
wavelength selection, and such devices are much slower to change state
than bespoke Pockels effect cells and drivers.  The prototype
instrument switches polarisation state at~\SI{5}{\hertz}, consisting
of a~\SI{50}{\milli\second} stabilisation period followed
by~\SI{50}{\milli\second} exposure time per polarisation phase, a
total of~\SI{200}{\milli\second} per observation data point.
Table~\ref{table:2018_performance} shows the noise performance for
five BiSON spectrophotometers over a 2018 summer observing campaign.
Mark-I at Iza{\~n}a switches at~\SI{0.5}{\hertz}.  Three instruments
at Las~Campanas in Chile, Narrabri in Australia, and Sutherland in
South Africa, all switch at~\SI{90.9}{\hertz}.  Comparing the faster
switching instruments with the slow-switching Mark-I instrument, we
see reduction in white noise power of~\SIrange{-6.6}{-11.5}{\decibel}
which as we have seen here is of the order expected through reduction
in scintillation noise alone.  The prototype instrument produces
mid-range performance inline with the faster switching instruments,
and is consistent with the scintillation noise profiles and
variability we have shown here.

Having such low noise is essential to allow the detection of very-low
frequency solar p-mode oscillations, which have amplitudes of a
centimetre-per-second or less~\citep{2014MNRAS.439.2025D}.


%% file: opendata.tex
%
%
%
%


\section{Open Data}

All code and data are freely available for download from the
University of Birmingham eData archive~\citep{edata412}, and also from
the source GitLab repository~\citep{gitlab}.


%% file: acknowledgements.tex
%
%
%


The authors are grateful for the financial support of the Science and
Technology Facilities Council (STFC), grant reference ST/R000417/1.
The BiSON node on the island of Tenerife is based at the Observatorio
del Teide operated by IAC, and we give particular thanks to Antonio
Pimienta and the team of operators who have contributed to running the
{Mark~I} instrument over many years.  Telescope operator logs provided
courtesy of Alex Oscoz, Head of Telescope Operations.  Dust data was
obtained with the STELLA robotic telescopes at Tenerife, an AIP
facility jointly operated by AIP and IAC.


%% file: paper.bbl
\begin{thebibliography}{18}
\providecommand{\natexlab}[1]{#1}
\providecommand{\url}[1]{\texttt{#1}}
\expandafter\ifx\csname urlstyle\endcsname\relax
  \providecommand{\doi}[1]{doi: #1}\else
  \providecommand{\doi}{doi: \begingroup \urlstyle{rm}\Url}\fi

\bibitem[{Davies} et~al.(2014){Davies}, {Broomhall}, {Chaplin}, {Elsworth}, and
  {Hale}]{2014MNRAS.439.2025D}
G.~R. {Davies}, A.~M. {Broomhall}, W.~J. {Chaplin}, Y.~{Elsworth}, and S.~J.
  {Hale}.
\newblock {Low-frequency, low-degree solar p-mode properties from 22 years of
  Birmingham Solar Oscillations Network data}.
\newblock \emph{\mnras}, 439:\penalty0 2025--2032, April 2014.
\newblock \doi{10.1093/mnras/stu080}.

\bibitem[{Dravins} et~al.(1997){Dravins}, {Lindegren}, {Mezey}, and
  {Young}]{1997PASP..109..173D}
D.~{Dravins}, L.~{Lindegren}, E.~{Mezey}, and A.~T. {Young}.
\newblock {Atmospheric Intensity Scintillation of Stars, I. Statistical
  Distributions and Temporal Properties}.
\newblock \emph{\pasp}, 109:\penalty0 173--207, Feb 1997.
\newblock \doi{10.1086/133872}.

\bibitem[{Dravins} et~al.(1998){Dravins}, {Lindegren}, {Mezey}, and
  {Young}]{1998PASP..110..610D}
D.~{Dravins}, L.~{Lindegren}, E.~{Mezey}, and A.~T. {Young}.
\newblock {Atmospheric Intensity Scintillation of Stars. III. Effects for
  Different Telescope Apertures}.
\newblock \emph{\pasp}, 110:\penalty0 610--633, May 1998.
\newblock \doi{10.1086/316161}.

\bibitem[{F{\"o}hring} et~al.(2019){F{\"o}hring}, {Wilson}, {Osborn}, and
  {Dhillon}]{2019MNRAS.489.5098F}
D.~{F{\"o}hring}, R.~W. {Wilson}, J.~{Osborn}, and V.~S. {Dhillon}.
\newblock {Atmospheric scintillation noise in ground-based exoplanet
  photometry}.
\newblock \emph{\mnras}, 489\penalty0 (4):\penalty0 5098--5108, Nov 2019.
\newblock \doi{10.1093/mnras/stz2444}.

\bibitem[{Hale}(2019{\natexlab{a}})]{edata412}
S.~J {Hale}.
\newblock {Measurement of atmospheric scintillation at Observatorio del Teide,
  I{\~z}ana, Tenerife}, December 2019{\natexlab{a}}.
\newblock URL \url{https://edata.bham.ac.uk/412/}.

\bibitem[{Hale}(2019{\natexlab{b}})]{gitlab}
S.~J {Hale}.
\newblock {Measurement of atmospheric scintillation at Observatorio del Teide,
  I{\~z}ana, Tenerife}, December 2019{\natexlab{b}}.
\newblock URL
  \url{https://gitlab.com/drstevenhale/izana-scintillation/-/tags/v1.0}.

\bibitem[{Hale}(2019{\natexlab{c}})]{halephd}
S.~J {Hale}.
\newblock \emph{Birmingham Solar Oscillations Network: The Next Generation}.
\newblock PhD thesis, School of Physics and Astronomy, University of
  Birmingham, UK, July 2019{\natexlab{c}}.
\newblock URL \url{https://etheses.bham.ac.uk/id/eprint/9010/}.

\bibitem[Hale et~al.(2016)Hale, Howe, Chaplin, Davies, and
  Elsworth]{s11207-015-0810-0}
S.~J. Hale, R.~Howe, W.~J. Chaplin, G.~R. Davies, and Y.~P. Elsworth.
\newblock {{Performance of the Birmingham Solar-Oscillations Network (BiSON)}}.
\newblock \emph{\solphys}, 291\penalty0 (1):\penalty0 1--28, January 2016.
\newblock ISSN 1573-093X.
\newblock \doi{10.1007/s11207-015-0810-0}.
\newblock URL
  \url{https://link.springer.com/article/10.1007%2Fs11207-015-0810-0}.

\bibitem[Hale et~al.(2017)Hale, Chaplin, Davies, Elsworth, Howe, Lund, Moxon,
  Thomas, Pall\'e, and {Rhodes}]{1538-3881-154-3-89}
S.~J. Hale, W.~J. Chaplin, G.~R. Davies, Y.~P. Elsworth, R.~Howe, M.~N. Lund,
  E.~Z. Moxon, A.~Thomas, P.~L. Pall\'e, and E.~J. {Rhodes}, Jr.
\newblock Atmospheric extinction coefficients in the $\mathrm{I_c}$ band for
  several major international observatories: Results from the {BiSON}
  telescopes, 1984-2016.
\newblock \emph{The Astronomical Journal}, 154\penalty0 (3):\penalty0 89, 2017.
\newblock \doi{10.3847/1538-3881/aa81d0}.
\newblock URL
  \url{https://iopscience.iop.org/article/10.3847/1538-3881/aa81d0}.

\bibitem[{Kenyon} et~al.(2006){Kenyon}, {Lawrence}, {Ashley}, {Storey},
  {Tokovinin}, and {Fossat}]{2006PASP..118..924K}
S.~L. {Kenyon}, J.~S. {Lawrence}, M.~C.~B. {Ashley}, J.~W.~V. {Storey},
  A.~{Tokovinin}, and E.~{Fossat}.
\newblock {Atmospheric Scintillation at Dome C, Antarctica: Implications for
  Photometry and Astrometry}.
\newblock \emph{\pasp}, 118:\penalty0 924--932, June 2006.
\newblock \doi{10.1086/505409}.

\bibitem[{Kornilov} et~al.(2012){Kornilov}, {Sarazin}, {Tokovinin},
  {Travouillon}, and {Voziakova}]{2012A&A...546A..41K}
V.~{Kornilov}, M.~{Sarazin}, A.~{Tokovinin}, T.~{Travouillon}, and
  O.~{Voziakova}.
\newblock {Comparison of the scintillation noise above different observatories
  measured with MASS instruments}.
\newblock \emph{\aap}, 546:\penalty0 A41, October 2012.
\newblock \doi{10.1051/0004-6361/201219954}.

\bibitem[{Osborn} et~al.(2015){Osborn}, {F{\"o}hring}, {Dhillon}, and
  {Wilson}]{2015MNRAS.452.1707O}
J.~{Osborn}, D.~{F{\"o}hring}, V.~S. {Dhillon}, and R.~W. {Wilson}.
\newblock {Atmospheric scintillation in astronomical photometry}.
\newblock \emph{\mnras}, 452:\penalty0 1707--1716, September 2015.
\newblock \doi{10.1093/mnras/stv1400}.

\bibitem[{Roca Cort{\'e}s} and {Pall{\'e}}(2014)]{2014MNRAS.443.1837R}
T.~{Roca Cort{\'e}s} and P.~L. {Pall{\'e}}.
\newblock {The Mark-I helioseismic experiment - I. Measurements of the solar
  gravitational redshift (1976-2013)}.
\newblock \emph{\mnras}, 443:\penalty0 1837--1848, September 2014.
\newblock \doi{10.1093/mnras/stu1238}.
\newblock URL \url{http://adsabs.harvard.edu/abs/2014MNRAS.443.1837R}.

\bibitem[Shen et~al.(2014)Shen, Yu, and Fan]{SHEN2014160}
Hong Shen, Longkun Yu, and Chengyu Fan.
\newblock Temporal spectrum of atmospheric scintillation and the effects of
  aperture averaging and time averaging.
\newblock \emph{Optics Communications}, 330:\penalty0 160 -- 164, 2014.
\newblock ISSN 0030-4018.
\newblock \doi{10.1016/j.optcom.2014.05.039}.
\newblock URL
  \url{http://www.sciencedirect.com/science/article/pii/S0030401814004969}.

\bibitem[{SIGRIST Photometer}(2019)]{14289E/4}
{SIGRIST Photometer}.
\newblock {VisGuard 2 In-situ Visibility Monitor}.
\newblock Technical Report 14289E/4, SIGRIST Photometer AG, Hofurlistrasse 1,
  6373 {Ennetb{\"u}rgen}, Switzerland, 2019.
\newblock URL
  \url{https://classic.photometer.com/svc/document.axd?id=14761&hl=E}.

\bibitem[{STELLA robotic telescopes}(2019)]{dust}
{STELLA robotic telescopes}.
\newblock {Environmental Status (Dust)}, 2019.
\newblock URL \url{http://stella.aip.de/stella/status/getdetail.php?typ=24}.

\bibitem[{Young}(1967)]{1967AJ.....72..747Y}
A.~T. {Young}.
\newblock {Photometric error analysis. VI. Confirmation of Reiger's theory of
  scintillation}.
\newblock \emph{\aj}, 72:\penalty0 747, August 1967.
\newblock \doi{10.1086/110303}.

\bibitem[{Young}(1969)]{1969ApOpt...8..869Y}
A.~T. {Young}.
\newblock {Photometric error analysis. VIII. The temporal power spectrum of
  scintillation.}
\newblock \emph{\ao}, 8:\penalty0 869--885, 1969.
\newblock \doi{10.1364/AO.8.000869}.

\end{thebibliography}
